\documentstyle[12pt]{article}
\begin{document}

\def\beq{\begin{equation}}
\def\eeq{\end{equation}}
\def\bea{\begin{eqnarray}}
\def\eea{\end{eqnarray}}
\def\ve{\vert}
\def\vel{\left|}
\def\ver{\right|}
\def\nnb{\nonumber}
\def\ga{\left(}
\def\dr{\right)}
\def\aga{\left\{}
\def\adr{\right\}}
\def\rar{\rightarrow}
\def\nnb{\nonumber}
\def\la{\langle}
\def\ra{\rangle}
\def\lla{\left<}
\def\rra{\right>}  
\def\ba{\begin{array}}
\def\ea{\end{array}}
\def\ds{\displaystyle}


\title{ {\Large {\bf
Distribution function of electrons and phonons in semiconductors and
semimetals in high electric and quantizing magnetic fields} }}

\author{\vspace{1cm}\\
{\large T. M. Gassym$^{a,b,}$\footnote{e-mail:~
gassymt@newton.physics.metu.edu.tr} } \\
\hskip 0.4 cm
$^a${\small Institute of Physics, National Academy of Sciences of
Azerbaijan Baku 370143, Azerbaijan}\\
\hskip 0.5 cm
$^b${\small Physics Department, Middle East Technical University
06531 Ankara, Turkey} }

\date{}

\begin{titlepage}
\maketitle
\thispagestyle{empty}

\begin{abstract}

The distribution function of electrons and phonons interacting with 
electrons in semiconductors and semimetals in high electric and quantizing 
magnetic fields as a result of the solution of the coupled system of 
equations for the density matrixes of electrons and phonons is obtained. 
The effects of heating of electrons and phonons and their arbitrary mutual 
drag are taken into account. The dispersion relation of energy of electrons 
is assumed to be arbitrarily spherically symmetric one. The spectrum of 
phonons is assumed to be isotropic.

The distribution function of electrons, phonons, the amplification
coefficient of phonons and the dependence of chemical potential on $E$,
$H$, $n$ and $T_e$ are obtained.

The distribution function of phonons is obtained for arbitrary drift
velocities of phonons. 
\end{abstract}
\end{titlepage}

\section{Introduction}
In the present paper, the behavior of semiconductors and semimetals in 
crossed high external electric ${\bf E}$ ($E=E_x$) and magnetic ${\bf H}$
($H=H_z$) fields is considered. The magnetic field is assumed to be high 
such that the cyclotron frequency $\Omega$ is much greater than the 
momentum relaxation frequency of electrons $\nu$, i.e. $\Omega \ll \nu$. 
It is known that an electron has stationary states at high electric and
magnetic fields. In Landau representation, the stationary states of
electrons are characterized by the magnetic quantum number $N$, the 
projection of momentum on the magnetic field direction $p_z$ and the
rotation center of electrons $X$. 

If the spectrum of electrons is assumed to be isotropic and quadratic,
then the eigenvalues of energy have the form\cite{1}:

\beq
\varepsilon_{\alpha}=\varepsilon_{N,p_z,X}=\hbar\Omega\left(N+\frac{1}{2} 
\right)+\frac{p_{z \alpha}^2}{2 m_n}-e E X_{\alpha}+
\frac{m_n \upsilon_{y \alpha}^2}{2}.
\eeq

The stationary state of electrons with energy given by Eq. (1) is 
characterized with Hall drift along the $y$ direction with velocity 
$\upsilon_{yo} \equiv V_H ={\ds \frac{cE}{H}}$. In the absence of scattering, 
the average velocity of electrons along the $x$ direction is equal to zero. 
The inclusion of scattering leads to the appearance of the conductance
current. The motion of electrons along the $x$ direction changes the
equilibrium position of the center of oscillation $X_0$, and, as a result,
changes the $p_y$ component of the electron momentum connected with them
$X_0=-{\ds \frac{c p_y}{e H}}$. In the case of scattering of electrons by
phonons, the motion of electrons along the $x$ direction is connected
with the transfer of the $y$ component of the momentum of electron $p_y$
to phonons, and leads to the stream of phonons along the $x$ axis, i.e., it 
leads to the mutual drag of electrons and phonons. During the motion
along the $x$ axis, the electron gains energy from the field, 
$eE(X_{\alpha}-X_{\beta})=e E X_{\alpha \beta} \equiv {\ds \frac{eER^2 q_y}
{\hbar}}$, and makes a transition from state $\alpha$ to state $\beta$. If
this energy is more than the emitted phonon energy $\hbar \omega_q$,
electrons and phonons are heated. Thus, the presence of external electric
field leads to the mutual drag of electrons and phonons. If the electric
field is high, then we have heating of electrons and phonons and their
mutual drag. 

In the quantized magnetic field the scattering frequencies of both electrons 
by phonons $\nu_p$ and phonons by electrons $\beta_e$ are increased, 
${\ds \left(\nu_p,~\beta_e \sim H^2\right)}$. On the other hand, the
interelectronic collision frequency $\nu_{ee}$ is decreased sharply. 
Because, for nondegenerate statistics of electrons only the lowest level of 
energy is fully occupied, and the number of electrons in higher energy levels 
is exponentially small. In this case the collision frequency between
electrons in different Landau levels becomes exponentially small. The
collisions between electrons in the lowest Landau level becomes elastic as
a result of the one dimensionality of the motion, and do not contribute to
the interelectronic relaxation. Then, we have ${\ds \frac{\nu_{ee}}{\nu_p}
\sim \left(\frac{T_e}{\hbar\Omega}\right)^2 exp\left(-\frac{\hbar\Omega}
{T_e}\right)}$. The decrease of interelectronic collisions in quantized
magnetic fields leads to a sharp decrease of the efficiency of redistribution 
of electrons in quantum states. Therefore, at high ${\bf E} \perp {\bf H}$
fields under the reasonable concentrations of electrons, the
approximation of ``effective electron temperature" is not satisfied. That
is a reason of why for the definition of distribution functions of electrons 
and phonons it is necessary to solve the coupled system of equations for
the density matrix of electrons and phonons directly.

Such a problem was formulated and solved for the first time in \cite{2} in 
the equilibrium state of phonons at lattice temperature $T$. This situation 
can be realized at high lattice temperature $T$, when the phonon--phonon 
collision frequency $\beta_p$ is much higher than the collision frequency 
of phonons by electrons $\beta_e$. However, for the quantization of orbital 
motion it is necessary to carry out the experiments at low temperatures 
(liquid Helium, or lower) of lattice when $\beta_p \rightarrow 0$. As it 
is shown experimentally\cite{3}--\cite{6}, under these conditions it is 
necessary to take into account the heating and mutual drag of electrons and 
phonons, and the generation of phonons by hot electrons. In the experimental
conditions the results of \cite{2} satisfy for drift velocities $V \ll s$. 
Nevertheless, as it follows from the result of \cite{2}, under the condition 
$V \ll s$, the heating of electrons is negligible. Moreover, in \cite{2} the 
change of chemical potential of electrons at high electric and magnetic 
fields is neglected. In fact, as it is shown in the present paper, for the 
nondegenerate statistics of electrons, the presence of high electric field
leads to the renormalization of chemical potential.

The present paper is devoted to solving the coupled system of equations for 
the diagonal parts of electrons and phonons density matrixes at high external 
crossed ${\bf E}$ and ${\bf H}$ fields with taking into account the heating 
of electrons and phonons, and their mutual drag. The problem is solved for 
arbitrary spherically symmetric spectrum of electrons. The spectrum of 
phonons is assumed to be isotropic.

\section{The spectrum of electrons in high electric and quantizing 
magnetic fields}
Let us assume that in the absence of electric field, the dispersion
relation of electrons is 

\beq
B(\varepsilon_{\alpha 0})=\frac{p_z^2}{2m_n}+\varepsilon_N,~~~~~~
\varepsilon_N=\hbar \Omega \left(N+\frac{1}{2}\right).  
\eeq

The energy of electrons $\varepsilon_{\alpha}$ at ${\bf E} \perp {\bf H}$ 
fields may also be written as

\beq
\varepsilon_{\alpha}=\varepsilon_{\alpha 0}-eEX_{\alpha}+
\frac{m(\varepsilon_{\alpha})\upsilon_{y\alpha}^2}{2}=\varepsilon_{\alpha
0}-eEX_{\alpha}+\frac{m(\varepsilon_{\alpha})}{m_n}\frac{m_n c^2}{2}
\frac{E^2}{H^2},
\eeq
where $m_n$ is the effective mass of electrons at the bottom of the
conduction band, and ${\ds m(\varepsilon_{\alpha})=m_n\left(
\frac{\partial B(\varepsilon_{\alpha})}{\partial \varepsilon_{\alpha}}
\right)}$ is the effective mass of electron. In the absence of the electric 
field,

\beq
X_{\alpha}=-\frac{p_{y\alpha}}{m_n\Omega}. 
\eeq

For the determination of the dispersion relation of electrons in ${\bf E}
\perp {\bf H}$ fields, we make a transition to the reference frame which
drifts together with electrons with a velocity of ${\ds V=\frac{cE}{H}}$. 
It is obvious that in such reference frame, the dispersion relation of
electrons must have the form of Eq. (2), and all properties of the system
must be preserved if we substitute

\beq
p_{y\alpha}^{\prime}=p_{y\alpha}+m(\varepsilon_{\alpha})V.  
\eeq
Then, by using Eq. (5) in Eq. (4), we may obtain $X_{\alpha}$ as

\beq
X_{\alpha}=-\frac{p_{y\alpha}}{m_n\Omega}-\frac{m(\varepsilon_{\alpha})V}
{m_n\Omega}=-\frac{p_{y\alpha}}{m_n\Omega}-\frac{eE}{m_n\Omega^2}
\left(\frac{\partial B(\varepsilon_{\alpha})}{\partial 
\varepsilon_{\alpha}}\right). 
\eeq

If we define $\varepsilon_{\alpha 0}$ as in Eq. (3) and substitute it into 
Eq. (2), we may obtain the dispersion relation of electrons in ${\bf E}
\perp {\bf H}$ fields as

\beq
B(\varepsilon_{\alpha}^{\star}) \equiv B \left(\varepsilon_{\alpha}
+eEX_{\alpha}-\frac{m(\varepsilon_{\alpha})}{m_n}\frac{m_n c^2}{2}
\frac{E^2}{H^2}\right)=\frac{p_{z\alpha}^2}{2m_n}+\hbar \Omega
\left(N+\frac{1}{2}\right),
\eeq
or,
\beq
\frac{p_{z\alpha}^2}{2m_n}=B(\varepsilon_{\alpha}^{\star})-\varepsilon_N.  
\eeq
$X_{\alpha}$ in Eq. (7) is determined from Eq. (6).

If the increasing of energy of electrons in electric field ${\ds\left(eE
X_{\alpha}-\frac{m(\varepsilon_{\alpha})c^2}{2}\frac{E^2}{H^2}\right)}$, 
is much less than $\varepsilon_{\alpha}$, by expanding 
$B(\varepsilon_{\alpha}^{\star})$ into a series around this small
parameter, we find

\beq
\frac{p_{z\alpha}^2}{2m_n} = B(\varepsilon_{\alpha})-\varepsilon_N+
\left[ eEX_{\alpha}-\frac{m_n c^2}{2}\left(\frac{\partial B(\varepsilon)}
{\partial \varepsilon_{\alpha}}\right)\right]\left(\frac{\partial 
B(\varepsilon_{\alpha})}{\partial \varepsilon_{\alpha}}\right),
\eeq
or,
\beq
\frac{p_{z\alpha}^2}{2m_n}=B(\varepsilon_{\alpha})-\varepsilon_N+
\frac{m(\varepsilon_{\alpha})}{m_n} e E X_{\alpha}-\frac{
m(\varepsilon_{\alpha})}{m_n}\frac{m_n c^2}{2}\frac{E^2}{H^2}.
\eeq

In the condition $V={\ds \frac{cE}{H} \ll \upsilon_{cr}=s}$, Eq. (10) 
reduces to the relations obtained earlier in \cite{7} and \cite{8} as a 
result of the solution of Schrodinger's equation for the Kane spectrum 
of electrons. In the present paper, Eq. (8) is obtained for arbitrary 
magnitude of the external electric field without any limit, and that is 
why this expression describes a more general case than the expressions 
given in \cite{7} and \cite{8}.

\section{The main equations and their solutions}
At high magnetic field ($\Omega \gg \nu$) in the Landau representation, 
the diagonal elements of the density matrix of electrons $f_{\alpha}$ is
larger than the nondiagonal elements ${\ds \left(\frac{\Omega}{\nu}
\right)}$ times, and that is why it is enough to write and solve the
equation for the diagonal elements of the density matrix of electrons.

Since in the present paper we consider that the space is uniform and the
temperature and concentration gradients are absent, it is enough to write 
the equations for the diagonal elements of density matrix $N({\bf q},t)$
for the phonon system.

After averaging over the electron states with fixed energies, the system of 
equations for the diagonal elements of the density matrixes of electrons
$f_{\alpha}=f(\varepsilon_{\alpha},t)$, and phonons $N_{{\bf q}}(t) \equiv 
N({\bf q},t)$ have the form\cite{9}:

\newpage
\bea
\frac{\partial f(\varepsilon,t)}{\partial t}=\frac{2\pi}{\hbar}
\sum_{\alpha,\beta,{\bf q}} \mid C_{{\bf q}} \mid^2 \vel \lla \alpha \mid
exp(-i{\bf q.r})\mid \beta \rra \ver^2\left\{\delta(\varepsilon_{\beta}-
\varepsilon_{\alpha}-\hbar\omega_q^{\star})[(f_{\beta}-f_{\alpha}) 
N({\bf q},t)+ \right. \\   
\left. \nonumber f_{\beta}(1-f_{\alpha})]-\delta(\varepsilon_{\beta}-
\varepsilon_{\alpha}+\hbar\omega_q^{\star})[(f_{\alpha}-f_{\beta})
N({\bf q},t)+f_{\alpha}(1-f_{\beta})] \right\}\delta(\varepsilon_{\alpha}
-\varepsilon)+I_{ee}[f]+I_{ed}[f],
\eea

\bea
\frac{\partial N({\bf q},t)}{\partial t}= \frac{4\pi}{\hbar}
\sum_{\alpha,\beta,{\bf q}}\mid C_{{\bf q}}\mid^2 \vel \lla \alpha \mid 
\exp(-i{\bf q}{\bf r}) \mid \beta \rra \ver^2 \delta(\varepsilon_{\beta}
-\varepsilon_{\alpha}-\hbar \omega_q^{\star}) \\
\nonumber
\left[(f_{\beta}-f_{\alpha}) N({\bf q},t)+f_{\beta}
(1-f_{\alpha})\right]+I_{pp}[N({\bf q})]+I_{pb}[N({\bf q})],
\eea
where ${\ds f(\varepsilon,t)=\sum_{\alpha}f(\varepsilon_{\alpha},t)
\delta (\varepsilon_{\alpha}-\varepsilon)}$ is the average number of 
carriers with energy $\varepsilon$, e.g., their distribution function,
${\ds \hbar \omega_q^{\star}=\hbar \omega_q-eE(X_{\beta}-X_{\alpha})}$, 
$I_{ee}[f]$ and $I_{ed}[f]$ are the inter--electronic and electron--defects 
collision integrals, respectively,
 
\beq
I_{pp}[N_{\bf q}]=\beta(q)[N({\bf q},t)-N({\bf q},T_p)],~~~~~~
I_{pb}[N_{\bf q}]=\beta_b[N({\bf q},t)-N({\bf q},T)],  
\eeq
are the phonon--phonon and phonon--crystal boundries collision integrals,
$T_p$ is the temperature of the heated phonons, and $T$ is the lattice 
temperature. Both phonon collision integrals are considered in the 
relaxation time approximation: ${\ds \beta_p^{-1}=\tau_p(q)}$, ${\ds 
\beta_b^{-1}=\tau_b}$\cite{10}. 

We consider the low electron concentration case $n< n_{cr}$ when $I_{ee} 
\ll I_{ep}$, where $I_{ep}$ is the collisions integral of electrons with 
phonons. For simplicity we neglect the contribution of collisions of
electrons with defects. Since in the Born approximation under quantizing 
magnetic field the collision frequency of electrons with neutral and 
ionized defects do not depend on the electron energy, the role of the 
scattering of electrons by defects may be easily taken into account in 
the final expressions.

Replacing $\alpha$ by $\beta$ in the second component of the expression 
under the sum in Eq. (11), we find

\bea
\frac{\partial f(\varepsilon,t)}{\partial t}=\frac{2\pi}{\hbar}
\sum_{\alpha,\beta,{\bf q}}\mid C_{{\bf q}}\mid^2 \vel \lla 
\alpha \mid \exp(-i{\bf q.r}) \mid \beta \rra \ver^2 \\ 
\nonumber
\left\{\delta(\varepsilon_{\beta}-\varepsilon_{\alpha}-\hbar 
\omega_q^{\star})[(f_{\beta}-f_{\alpha}) N({\bf q},t)+f_{\beta}
\left(1-f_{\alpha}\right)]\right\}\left\{\delta(\varepsilon_{\alpha}
-\varepsilon)-\delta(\varepsilon_{\beta}-\varepsilon)\right\}.  
\eea

We consider the case when the scattering of electrons by phonons is 
quasielastic and, therefore, changing of the energy of electrons 
$\varepsilon_{\alpha}-\varepsilon_{\beta}=\hbar\omega_q-Vq_y=\hbar 
\omega_q^{\star}$ is less than the energy scale of the changing electrons 
distribution function.

Then, expanding $(f_{\beta}-f_{\alpha})$ and $[\delta(\varepsilon_{\alpha}
-\varepsilon)-\delta(\varepsilon_{\beta}-\varepsilon)]$ into series, we find
\beq
f_{\beta}-f_{\alpha}=f(\varepsilon_{\alpha}+\hbar\omega_q^{\star})-
f(\varepsilon_{\alpha}) \approx \hbar\omega_q^{\star}\left(\frac{\partial 
f(\varepsilon_{\alpha})}{\partial \varepsilon_{\alpha}}\right)+....,
\eeq
\beq
\left\{\delta(\varepsilon_{\alpha}-\varepsilon)-\delta(\varepsilon_{\beta}
-\varepsilon)\right\}=\delta(\varepsilon_{\alpha}-\varepsilon)-\delta
(\varepsilon_{\alpha}-\varepsilon+\hbar\omega_q^{\star})\approx-\hbar 
\omega_q^{\star}~\frac{\partial}{\partial \varepsilon_{\alpha}}~
\delta (\varepsilon_{\alpha}-\varepsilon)+..., 
\eeq

By using the identity ${\ds \frac{\partial}{\partial \varepsilon_{\alpha}}
\delta(\varepsilon_{\alpha}-\varepsilon)\equiv\frac{\partial}{\partial 
\varepsilon}\delta(\varepsilon_{\alpha}-\varepsilon)}$, we obtain

\beq
{\ds \frac{\partial f(\varepsilon,t)}{\partial t}=-\frac{\partial}
{\partial\varepsilon}\left\{ A(\varepsilon)\frac{\partial 
f(\varepsilon,t)}{\partial \varepsilon}+D(\varepsilon)f(\varepsilon,t)
[1-f(\varepsilon,t)]\right\}},
\eeq
where, 
\bea
A(\varepsilon)=\frac{2\pi}{\hbar}\sum_{\alpha,\beta,{\bf q}}\mid C_{{\bf q}} 
\mid^2~ \mid I_{\alpha\beta} \mid ^2(\hbar\omega_q^{\star})^2 \delta
\left(\varepsilon_{\beta}-\varepsilon_{\alpha}-\hbar\omega_q^{\star}\right) 
N({\bf q},t)\delta\left(\varepsilon_{\alpha}-\varepsilon \right), \\
\nonumber
D(\varepsilon)=\frac{2\pi}{\hbar}\sum_{\alpha,\beta,{\bf q}}\mid C_{{\bf q}}
\mid^2 \mid I_{\alpha\beta}\mid^2 \hbar \omega_q^{\star}~\delta
\left(\varepsilon_{\alpha}-\varepsilon_{\beta}-\hbar\omega_q^{\star}
\right)\delta\left(\varepsilon_{\alpha}-\varepsilon \right).
\eea
The stationary solution of Eq. (17) satisfying boundary condition 
$\lim_{\varepsilon \rightarrow \infty} f(\varepsilon) \rightarrow 0$ is

\beq
f(\varepsilon)=\left\{const.^{-1}\exp\left(\int^{\varepsilon}\frac{
d\varepsilon^{\prime}}{T_e(\varepsilon^{\prime})}\right)+1\right\}^{-1},
\eeq
where $T_e(\varepsilon)={\ds \frac{A(\varepsilon)}{D(\varepsilon)}}$ is
the temperature of electrons with energy $\varepsilon$.

The solution of Eq. (12) is

\beq
N({\bf q},t)=\left\{N(q,0)+\beta\gamma_q^{-1}~\widetilde{N}(q)\right\}
\exp \left(\gamma_q t \right)-\beta\gamma_q^{-1}~\widetilde{N}(q),
\eeq
where ${\ds \gamma_q=\beta\left({\ds \frac{{\bf u.q}}{\hbar\omega_q}}-1
\right)}$ is the increment of the generation of phonons, $\beta=\beta_e
+\beta_p+\beta_b$ is the total collision frequencies of phonons by the 
scatterers and

\bea
\beta_e=\frac{2\pi}{\hbar}\sum_{\alpha,\beta}\mid C_{\bf q} \mid^2 \mid 
I_{\alpha,\beta}\mid^2 \left(f_{\beta}-f_{\alpha}\right)\delta\left(
\varepsilon_{\beta}-\varepsilon_{\alpha}-\hbar\omega_q^{\star}\right) \\
\nonumber 
\approx \frac{2\pi}{\hbar}\sum_{\alpha,\beta,{\bf q}}\mid C_{{\bf q}}
\mid^2 \mid I_{\alpha\beta}\mid^2(\varepsilon_{\beta}-\varepsilon_{\alpha})
\left(\frac{\partial f(\varepsilon_{\alpha})}{\partial \varepsilon_{\alpha}}
\right)\delta\left(\varepsilon_{\beta}-\varepsilon_{\alpha}-\hbar 
\omega_q^{\star}\right).
\eea

$N(q,0) \equiv N(q,T)$ is the initial distribution function of phonons at 
$t=0$ in the absence of external fields,

\beq
u=\left(1-\frac{\beta_p}{\beta}\right)V,~~~~~~\mid I_{\alpha \beta} 
\mid^2=\vel \lla \alpha \mid exp(-i{\bf q.r}) \mid \beta \rra \ver^2.
\eeq

As it follows from Eq. (20) in the $\gamma_q > 0$ case, i.e., when the 
drift velocity of phonons $u$ is larger than the sound velocity $s$, the 
distribution function of phonons $N({\bf q},t)$ increases exponentially
with time, whereas in the $\gamma_q<0$ case the solution, Eq. (20), is 
stationary,

\beq
N({\bf q})=\lim_{t \rightarrow \infty} N({\bf q},t)=-\beta \gamma_q^{-1}
\widetilde{N}(q),~~~~~~\widetilde{N}(q)=\gamma_e N({\bf q},T_e)+
\gamma_p N(q,T_p),
\eeq
where $\gamma_e={\ds \frac{\beta_e}{\beta}}$,~$\gamma_p={\ds 
\frac{\beta_p}{\beta}}$,

\beq
N(q,T_e)=\frac{{\ds \sum_{\alpha,\beta}\mid I_{\alpha\beta} \mid^2 
T_e(\varepsilon)\left[\partial f(\varepsilon_{\alpha})/ \partial
\varepsilon_{\alpha}\right]\delta(\varepsilon_{\beta}-\varepsilon_{\alpha}
-\hbar \omega_q^{\star})}}{{\ds \sum_{\alpha\beta}\mid I_{\alpha\beta}
\mid^2 \hbar \omega_q^{\star}\left[\partial f(\varepsilon_{\alpha})/
\partial \varepsilon_{\alpha}\right]\delta(\varepsilon_{\beta}
-\varepsilon_{\alpha}-\hbar\omega_q^{\star})}},
\eeq

\beq
T_e=\frac{{\ds \sum_{\alpha,\beta}\mid I_{\alpha\beta} \mid^2
T_e(\varepsilon)\left[\partial f(\varepsilon_{\alpha})/
\partial\varepsilon_{\alpha}\right]\delta(\varepsilon_{\beta}
-\varepsilon_{\alpha}-\hbar \omega_q^{\star})}}{{\ds \sum_{\alpha\beta}
\mid I_{\alpha\beta}\mid^2 \left[\partial f(\varepsilon_{\alpha})/
\partial \varepsilon_{\alpha}\right]\delta(\varepsilon_{\beta}
-\varepsilon_{\alpha}-\hbar \omega_q^{\star})}}.
\eeq
where $T_e$ is the effective temperature of electrons. 

In the case of $\gamma_q=0$ or ${\ds \frac{\bf u.q}{\hbar\omega_q}
\approx 1}$ or $\hbar \omega_q^{\star}=0$, from Eq. (17) we have ${\ds
\frac{\partial f(\varepsilon,t)}{\partial t}=0}$, or $f(\varepsilon,t)
=const$. At this point the distribution function of electrons 
$f(\varepsilon,t)$, the effective electron temperature $T_e$ and the drift 
velocity of electrons are constant. Therefore, the state is nondissipative 
and the current is constant. Moreover, the distribution function of 
phonons is nonstationary and grows by time linearly, $N(q,t)=N(q,T)+\beta 
t N(q,T_e)$. Indeed, $u=V=s=const$. and ${\ds \frac{d N(q,t)}{dt}=\beta 
N(q,T_e)}=const$. Namely, ${\ds P(T_e)=\sum_q \hbar \omega_q^{\star}
\left(\frac{d N(q,t)}{d t}\right)=0}$, where $P(T_e)$ is the power 
transferred by electrons to phonons. The point $u=s$ is the acoustical
instability threshold (AIT). At this point, the stimulated emission of
phonons is equal to the stimulated absorption of phonons, and we have only 
spontaneous emission of phonons at high external electric and magnetic
fields. At this point collisions of electrons with phonons are exactly
elastic, i.e., state is nondissipative and dynamically stationary because
of the power received from electric field emitted as phonons by the
process of the stimulated emission. Moreover, at this point the current
$J=const$. and $v_d=V={\ds \frac{c E}{H}}=s$, or $cE=sH$, i.e., 
$E_{cr}={\ds \frac{sH}{c}}$ or $H_{cr}={\ds \frac{cE}{s}}$.

Substituting ${\ds \frac{\partial N({\bf q},T_e)}{\partial t}=0}$ in Eq. 
(12), we may directly solve this equation under the boundary conditions

\beq
N({\bf q},T_e) \mid_{t=0}=N({\bf q},0) \equiv N({\bf q},T)=
\left\{\exp \left(\frac{\hbar \omega_q}{T} \right)-1 \right\}^{-1},
\eeq
and we obtain Eq. (23).

Under the considered conditions $\hbar \omega_q^{\star} \ll T_e,T$ from 
Eq. (23), we find

\beq
\widetilde{N}({\bf q})=\frac{\gamma_e T_e+\gamma_p T_p}{\hbar\omega_q}
=\frac{\widetilde{T}}{\hbar \omega_q}=\widetilde{N}(q,\widetilde{T}), 
~~~~~~\widetilde{T}=\gamma_e T_e+\gamma_p T_p, 
\eeq
where $\widetilde{T}$ is the temperature of coupled by the mutual drag 
system of electrons and phonons. Therefore, the stationary solution of Eq. 
(12) has the form $(\hbar \omega_q^{\star} > 0)$:

\bea
N({\bf q})=\frac{\widetilde{N}({\bf q})}{1-{\ds {\frac{\bf u.q}{\hbar 
\omega_q}}}}=-\beta~\gamma_q^{-1}~\widetilde{N}(q) \approx 
\frac{\widetilde{T}}{\hbar\omega_q^{\star}}.  
\eea

Let us substitute Eq. (28) in Eq. (18) and take into account the relations:

\bea
\vel I_{\alpha \beta} \ver^2 =\vel I_{N N^{\prime}}\ver^2~ 
\delta_{p_{y\beta},p_{y}+q_y}~\delta_{p_{z\beta},p_{z}+q_z}, \\ 
\nonumber
\mid I_{N N^{\prime}}\mid^2=\left(\frac{N!}{N^{\prime}!}\right)^{1/2}
\exp \left(-\frac{q_{\perp}^2}{q_H^2}\right)\left(-\frac{q_{\perp}}{q_H}
\right)^{N^{\prime}-N} L^{\mid N^{\prime}-N \mid}\frac{q_{\perp}^2}{q_H^2}.  
\eea
Here ${\ds q_H=\hbar R^{-1}=(m_n \varepsilon_N)^{1/2}}$,~${\ds q_{\perp}^2
=q_x^2+q_y^2}$, ${\ds L_N^{\mid m \mid}}$ is the Laguerre's polynomial 
normalized to unity, and $\mid m \mid =\mid N-N^{\prime} \mid$. 

Choosing the cylindrical coordinate system with direction along the
magnetic field

\bea
dq=q_{\perp}~dq_{\perp}~dq_z~ d\varphi, ~~~~~~q_y=q_{\perp} \sin\varphi.
\eea
As a result of the integration, we have

\begin{eqnarray}
A(\varepsilon)=\widetilde{T} \Phi(u/s) D(\varepsilon ),~~~~~~
\Phi(u/s)=\varphi-2 \frac{V}{u}(\varphi_1-1)-\left(\frac{V}{u}
\right)^2 (\varphi_1-1)^2, \\
\nonumber 
D(\varepsilon)=\frac{s\Omega}{\hbar (2\pi \hbar)^5}\sum_{N,N^{\prime}} 
q_H^3 W(q_H)\int_0^{\infty} dx~\frac{x^2 \mid I_{N N^{\prime}}\mid^2~
m^2(\varepsilon)}{\left[B(\varepsilon+\hbar\omega_q^{\star})-
\varepsilon_{N^{\prime}}\right]^{1/2}\left[B(\varepsilon-\hbar 
\omega_q^{\star})-\varepsilon_N\right]^{1/2}},
\end{eqnarray}
where $x=q_{\perp}/q_H$, and $W(q_H)$ is the constant part of 
the potential in mutual interaction of electrons by phonons.

Thus, in general, under the arbitrary degree of quantization, the ratio of
$A(\varepsilon)$ to $D(\varepsilon)$ is $\widetilde{T}\Phi(u/s)$; and it 
does not depend on the energy of electrons and the potential of mutual 
interactions with phonons. In other words, in a more general case of 
interaction of electrons by the acoustic and optical phonons the expression:

\bea
T_{ef}=\frac{A(\varepsilon)}{D(\varepsilon)}=\widetilde{T}\Phi(u/s),  
\eea
does not depend on the energy of electrons.

By substituting Eq. (32) into Eq. (19), we find

\beq
f(\varepsilon)=\left\{1+\exp\left(\frac{\varepsilon-
\zeta(E,H)}{T_{ef}} \right) \right\}^{-1}.  
\eeq

In other words, at quantizing magnetic fields the distribution function of 
electrons is a Fermi one, and includes the effective temperature in the
following form:

\beq
T_{ef}=\widetilde{T}\left\{1+\left(1-\frac{V}{u}\right)^2(\varphi_1-1)
\right\}, ~~~~~~\varphi_1=\left(1-\frac{u^2}{s^2}\right)^{-1/2}.
\eeq

In the classical region of strong magnetic fields ($\Omega \gg \nu$)

\beq
\Phi(u/s)=\left\{1+\left(1-\frac{V}{u}\right)(\varphi_2-1)+\frac{1}{3}
\left(\frac{V}{s}\right)^2\right\},~~~~~~~~\varphi_2=\frac{s}{2u}\ln 
\vel \frac{s+u}{s-u} \ver.  
\eeq

As it follows from Eq. (33), the distribution function of electrons is 
the Fermi one with effective temperature:

\bea
T_{ef}=\widetilde{T}\left\{1+\left(1-\frac{V}{u}\right)(\varphi_2-1)
+\frac{1}{3}\left(\frac{V}{s}\right)^{2}\right\}.
\eea

The fact that the distribution function of electrons in both classical 
and quantum regions of magnetic fields are the Fermi ones with effective 
temperature is a result of the independency of drift velocity of electrons
$V={\ds \frac{cE}{H}}$ from energy of electrons $\varepsilon$.

At small values of the drift velocity of electrons, $V \ll s$, from
Eq. (34) for the quantizing magnetic field, we have

\bea
T_{ef} \approx \widetilde{T} \left\{1+\frac{1}{2} \left(\frac{V}{s}
\right)^2-\frac{u V}{s^2} \right\}=\widetilde{T}\left\{1+\left(\gamma_p 
-\frac{1}{2}\right)\left(\frac{c E}{s H}\right)^2 \right\}. 
\eea

If $\beta_e \gg \beta_p$, ${\ds T_{ef}=\widetilde{T}\left\{1-\frac{1}{2}
\left(\frac{c E}{s H}\right)^2\right\}}$; and if $\beta_e \ll \beta_p$,
${\ds T_{ef}=\widetilde{T}\left\{1+\frac{1}{2}\left(\frac{c E}
{s H}\right)^2\right\}}$.

In the classic region of magnetic fields

\beq
T_{ef}\approx \widetilde{T}\left\{1+\frac{1}{3}\left(\frac{V}{s}\right)^2
\right\}=\widetilde{T}\left\{1+\frac{1}{3}\left(\frac{cE}{sH}\right)^2
\right\}.  
\eeq

If $\beta_p \gg \beta_e$, phonons are not heated, $\widetilde{T}=T$ and 
$V \ll s$. Under these conditions from Eqs. (37) and (38) we can obtain 
the results of \cite{2}.

\section{The statistics of electrons at high magnetic fields}
In crossed ${\bf E} \perp {\bf H}$ field, the chemical potential of 
electrons may be obtained from the normalization condition of the
distribution function as

\beq
n_e=\int^{\infty}d\varepsilon~ g_H(\varepsilon) f(\varepsilon),~~~~~~
g_H(\varepsilon)=\frac{2m_n (eH/c)}{(2\pi\hbar)^2}\sum_N 
p_z^{-1}(\varepsilon).  
\eeq

For arbitrary spherical symmetric spectrum of electrons

\beq
n_e=\frac{2 (2m_n)^{1/2} m_n \Omega}{(2 \pi \hbar)^2}\sum_{N} 
\int_{\varepsilon_1^{\star}}^{\infty}d\varepsilon^{\star}~\left[
B(\varepsilon^{\star})-\varepsilon_N \right]^{1/2}\left\{1+\exp\left(
\frac{\varepsilon^{\star}-\zeta^{\star}(E,H)}{T_{ef}}\right)\right\}^{-1},
\eeq

\beq
\zeta^{\star}(E,H)=\zeta(E,H)+eEX(\varepsilon)-
\frac{m(\varepsilon)c^2}{2}\left(\frac{E^2}{H^2}\right).  
\eeq

The $\varepsilon_1^{\star}$ is determined as a solution 

\beq
B(\varepsilon^{\star})-\varepsilon_N=0.
\eeq
By partial integration of Eq. (40), we find

\beq
n_e=\frac{4(2m_n)^{1/2}m_n\Omega}{(2\pi\hbar)^2}\sum
\int_{\varepsilon_{1}^{\star}}^{\infty}d\varepsilon^{\star}
\left[B(\varepsilon^{\star})-\varepsilon_N\right]^{1/2}
\left(-\frac{\partial f \left(\varepsilon^{\star}\right)}
{\partial \varepsilon^{\star}}\right).
\eeq

For the case of parabolic spectrum of electrons

\beq
\varepsilon_1^{\star}=\varepsilon_N=\hbar\Omega\left(N+\frac{1}{2}\right),  
\eeq
and for the Kane spectrum of electrons:

\beq
\varepsilon^{\star}_1=-\frac{\varepsilon_g}{2} \left[1-
\left(1+\frac{4\varepsilon_N}{\varepsilon_g}\right)^{1/2}\right].
\eeq

Let us consider the case of degenerate and non--degenerate statistics of 
electrons alone. For strong degenerate electrons

\beq
-\left(\frac{\partial f_0(\varepsilon^{\star})}{\partial
\varepsilon^{\star}}\right)=\delta(\varepsilon-\zeta^{\star}). 
\eeq
With the help of this expression, we can integrate the Eq. (43) and get

\beq
n_e=\frac{4(2m_n)^{1/2}}{\hbar(2\pi R)^2}\sum_N\left[B^{\star}
(\zeta^{\star})-\hbar\Omega\left(N+\frac{1}{2}\right)\right]^{1/2},
\eeq

For the parabolic spectrum of electrons

\beq
n_e=\frac{4(2m_n)^{1/2}}{\hbar (2\pi R)^2}\sum_{N} 
\left[\zeta^{\star}-\hbar \Omega \left(N+\frac{1}{2}
\right)\right]^{1/2}. 
\eeq

For the Kane spectrum of electrons

\beq
n_e=\frac{4(2m_n)^{1/2}}{\hbar (2\pi R)^2}\sum_N\left[\zeta^{\star}
\left(1+\frac{\zeta^{\star}}{\varepsilon_g}\right)-\hbar\Omega
\left(N+\frac{1}{2}\right) \right]^{1/2}. 
\eeq

For ultraquantum limits ($N=N^{\prime}=0$), we have

\bea
n_e &=& \frac{4(2m_n)^{1/2}}{\hbar (2\pi R)^2}\left[B\left(
\zeta^{\star}\right)-\frac{\hbar\Omega}{2}\right]^{1/2} \\
\nonumber
&=& B(\zeta^{\star})-\frac{\hbar \Omega}{2}=
\frac{\hbar^2 (2 \pi R)^4 n_e^2}{16 (2m_n)},
\eea
where ${\ds B(\zeta^{\star})=\frac{\hbar\Omega}{2}+\frac{h^2(2\pi R)^4 
n_e^2}{16 (2 m n)}}$. From this relation for the parabolic spectrum of
electrons we obtain the chemical potential as follows:

\beq
\zeta^{\star}(E,H)=\frac{\hbar \Omega}{2}+\frac{\pi^4 \hbar^4 n_e^2}
{2m_n^3 \Omega^2}.
\eeq
This expression is the same of Eqs. (32) and (22) in \cite{12}. 

In the case of Kane spectrum of electrons:

\beq
\zeta^{\star}(E,H)=-\frac{\varepsilon_g}{2}\left\{1-\left(1+
\frac{2 \hbar \Omega}{\varepsilon_g}+\frac{2 \pi^4 \hbar^2 R^4 
n_e^2}{m_n \varepsilon_g}\right)^{1/2}\right\}. 
\eeq

In the case of nondegenerate electrons by taking $[B(\varepsilon^{\star})
-\varepsilon_N]$ out of the integral when $\varepsilon^{\star}=
\varepsilon_1^{\star}(N)+T_e$, we find

\beq
\exp \left(\frac{\zeta^{\star \star}(E,H)}{T_e}\right)=\frac{\left(2 
\pi R \right)^2 \hbar n_e}{4(2m_n)^{1/2}}\left\{\sum_N \exp\left(-
\frac{\varepsilon_1^{\star}(N)}{T_e}\right)\left[B(\varepsilon_1^{\star}
(N)+T_e)-\varepsilon_N\right]^{1/2}\right\}^{-1}. 
\eeq

As before $\varepsilon_1^{\star}$ is obtained from Eqs. (44) and (45). If 
the condition $\varepsilon_1^{\star} \gg T_e$ is satisfied, then by
expanding $B(\varepsilon_1^{\star}+T_e)$ in Eq. (53) into series and by
taking into account Eq. (42), we have

\beq
\exp \left(\frac{\zeta^{\star \star}(E,H)}{T_e}\right) \simeq 
\frac{\left(2\pi R\right)^2\hbar n_e}{4\sqrt{2}}~T_e^{-1/2}\left[\sum_N 
\exp \left(-\frac{\varepsilon_1^{\star}(N)}{T_e}\right) m^{1/2}
(\varepsilon_1^{\star})\right]^{-1},
\eeq
where
\beq
m(\varepsilon_1^{\star})=m_n \left(\frac{\partial 
B(\varepsilon^{\star})}{\partial \varepsilon^{\star}}
\right)_{\varepsilon^{\star}=\varepsilon_1^{\star}}.
\eeq

Dividing Eq. (53) by itself for $E=0$, we get

\beq
\zeta^{\star \star}(E,H)=\zeta(H)\frac{T_e}{T}+T_e \ln\left(\frac{n_e}
{n_0}\right)+T_e \ln\left[\frac{F_N(T)}{F_N(T_e)}\right],  
\eeq
where,
\beq
\zeta^{\star \star}(E,H)= \zeta(E,H)-\frac{m}{m_n}
\frac{m_n c^2}{2}\frac{E^2}{H^2}, 
\eeq

\beq
F_N(T_e)=\sum_N \exp \left(-\frac{\varepsilon_1^{\star}(N)}{T_e}\right)
\left\{B(\varepsilon_1^{\star}(N)+T_e)-\varepsilon_N \right\}^{1/2}. 
\eeq
$F_N(T)$ may be obtained from Eq. (58) by replacing $T_e$ with $T$.

For the case of parabolic spectrum of electrons ${\ds\varepsilon_1^{\star}
=\varepsilon_N=\hbar \Omega\left(N+\frac{1}{2}\right)}$, and, we have 

\beq
\zeta^{\star \star}(E,H) =\zeta(H) \frac{T_e}{T}+T_e 
\ln\left(\frac{n_e}{n_0}\right)-\frac{T_e}{2}
\ln\left(\frac{T_e}{T}\right)+ T_e \ln\left[\frac
{\sinh(\hbar \Omega/2 T_e)}
{\sinh(\hbar \Omega/2 T)}\right].
\eeq
Therefore, in the case of parabolic spectrum of electrons we finally
find the chemical potential as

\bea
\zeta(E,H) =\zeta(H) \frac{T_e}{T}-eEX+\frac{m_n c^2}{2}
\frac{E^2}{H^2}+T_e \ln\left(\frac{n_e}{n_0}\right)+ \\
\nonumber
\frac{T_e}{2} \ln\left(\frac{T_e}{T}\right)+T_e\ln\left(\frac{
\sinh(\hbar\Omega/2 T_e)}{\sinh(\hbar \Omega/2 T)}\right].
\eea

As it follows from Eq. (60) in external electric field if the 
concentration of electrons is increased ($n_e>n$), then the chemical
potential of electrons must also increase.

If we have full ionization of small impurity centers, then 
$n_e=n_0=const$. Thus, from Eq. (60), we may obtain

\bea
\zeta(E,H)-\frac{\hbar\Omega}{2}=\frac{T_e}{T}\left[\zeta(H)-
\frac{\hbar\Omega}{2}\right]+\frac{m_n c^2}{2}\frac{E^2}{H^2}- \\
\nonumber
\frac{T_e}{2} \ln\left(\frac{T_e}{T}\right)+T_e\ln \left[\frac{1-
\exp(\hbar \Omega /2 T_e)}{1-\exp(\hbar \Omega /2 T)}\right]. 
\eea

In the quantizing magnetic fields $\hbar \Omega > T_e,T$ and, for this 
reason, the last term in Eq. (61) may be presented as ${\ds \frac{T_e}
{T}\frac{\hbar \Omega}{2}}$. In this case

\beq
\zeta(E,H)-\frac{\hbar \Omega}{2}=\frac{T_e}{T}[\zeta(H)-\hbar\Omega]+
\frac{m_n c^2}{2}\frac{E^2}{H^2}-\frac{T_e}{2}\ln\left(\frac{T_e}{T}\right). 
\eeq

For the weak electric fields $T_e=T$,

\beq
\zeta(E,H)=\zeta(H)-\frac{\hbar\Omega}{2}+\frac{m_n c^2}{2}\frac{E^2}{H^2}. 
\eeq

As it follows from Eq. (63), the expression for $\zeta(E,H)$ differs from
the expression given in \cite{13} by the factor ${\ds \frac{m_n c^2}{2} 
\frac{E^2}{H^2}}$, which is connected with the Hall drift of electrons. 
In the case of heating of electrons at external electric field the main 
contribution to the free energy is obtained by the expression ${\ds \zeta(H)
\frac{T_e}{T}}$, i.e., in high external electric field the chemical potential 
of electrons in common case increases linearly by ${\ds \frac{T_e}{T}}$.

For the calculation of the statistical behavior of electrons in external 
fields such as magnetic susceptibility, heat capacity and {\it etc}., it
is necessary to know the dependence of the chemical potential on the
intensity of external electric and magnetic fields.

The magnetic susceptibility of hot electrons in the case of high 
concentration of electrons was investigated earlier\cite{14}.

\section{Conclusion}
In the present work, it is shown that under the conditions of arbitrary
degree of quantization and for the interaction of electrons with both the
acoustical and optical phonons, the distribution function of electrons has
the form of Fermi distribution function with effective electron temperature. 
This result is obtained in the case of small concentration of electrons
$n \leq n_{cr}$ when the usual approximation of ``effective temperature" 
for electrons is not satisfied, i.e., $\nu_{ee} \ll \nu_p$.

The distribution function of phonons interacting with electrons is obtained
for arbitrary drift velocities of phonons. If the drift velocity of phonons
$u$ is smaller than the sound velocity $s$ there is a stationary state and
the distribution function of phonons is shifted Planck's one with effective
temperature of phonons. In the case of $u>s$, the distribution function
of phonons grows with time exponentially, i.e., we have the effect of 
generation or amplification of phonons by electric field. The amplification
coefficient of phonons ${\ds \gamma_q=\frac{\beta}{s}\left(\frac{{\bf u.q}}
{\hbar \omega_q}-1\right)}$ is obtained. It has been established that in
the considered case the chemical potential of electrons is renormalised. 
The $E$, $H$, $n$ and $T_e$ dependences of the chemical potential are
obtained.

\section*{Acknowledgments}
This work was partially supported by the Scientific and Technical
Research Council of Turkey (TUBITAK). In the course of this work,
T. M. Gassym was supported by a TUBITAK--NATO fellowship.

\end{document}